# Knowledge Graphs in the Digital Twin – A Systematic Literature Review About the Combination of Semantic Technologies and Simulation in Industrial Automation

**Franz Georg Listl[1, 2], Daniel Dittler[1], Gary Hildebrandt[3], Valentin Stegmaier[1, 4], Nasser Jazdi[1], Member, IEEE, and Michael Weyrich[1]**

[1]Institute of Industrial Automation and Software Engineering, University of Stuttgart, Stuttgart, Germany
[2]Technology, Siemens AG, Munich, Germany
[3] Institute of Smart Systems and Services, Pforzheim University, Pforzheim, Germany
[4] J. Schmalz GmbH, Glatten, Germany

Corresponding author: Franz Georg Listl (e-mail: franz.listl@siemens.com).

**ABSTRACT** The ongoing digitization of the industrial sector has reached a pivotal juncture with the emergence of Digital Twins, offering a digital representation of physical assets and processes. One key aspect of those digital representations are simulation models, enabling a deeper insight in the assets current state and its characteristics. This paper asserts that the next evolutionary step in this digitization journey involves the integration of intelligent linkages between diverse simulation models within the Digital Twin framework. Crucially, for the Digital Twin to be a scalable and cost-effective solution, there is a pressing need for automated adaption, (re-)configuration, and generation of simulation models. Recognizing the inherent challenges in achieving such automation, this paper analyses the utilization of knowledge graphs as a potentially very suitable technological solution. Knowledge graphs, acting as interconnected and interrelated databases, provide a means of seamlessly integrating different data sources, facilitating the efficient integration and automated adaption of data and (simulation) models in the Digital Twin. We conducted a comprehensive literature review to analyze the current landscape of knowledge graphs in the context of Digital Twins with focus on simulation models. By addressing the challenges associated with scalability and maintenance, this research contributes to the effective adaption of Digital Twins in the industrial sector, paving the way for enhanced efficiency, adaptability, and resilience in the face of evolving technological landscapes.

## I. INTRODUCTION

In the landscape of industrial automation technology, Digital Twins emerge as a pivotal innovation, enabling the efficient implementation of various use cases of digital manufacturing such as remote monitoring, predictive analytics or simulating future behavior [1, 2]. The concept of Digital Twins has revolutionized the way production systems can be perceived, monitored and optimized [3]. Central to a Digital Twin are models, enabling an abstract representation of the asset from multiple perspectives, as well as the data describing different aspects of the asset itself [4]. However, including such models and data on a large scale also raises challenges including the need for semantic annotated data and models. To illustrate the existing need for semantically annotated data and models in Digital Twins, three exemplary use cases are presented below.

A first use case refers to behavior models, also referred to as simulation models in their executable form, of multiple components. They are provided by component manufacturers and used by machine manufacturers or plant operators [5]. However, manual intervention is often required to integrate such simulation models, as inputs and outputs as well as parameters and the correct linking of simulation model configurations are not directly available in machine-readable form. An expert with appropriate knowledge can link the inputs of a components simulation model directly to the corresponding signals in the overall system model. A software program, on the other hand, cannot create the corresponding links without this knowledge. A knowledge graph could fill this knowledge gap and enable automating this process [6]. An example of such a component model in a system model is shown in Fig. 1. It involves interpreting the inputs and outputs of the models as well as appropriately simulating the sequence for each production scenario, which cannot simply be linked to the signals in the system model without further information or expert knowledge. The pressure supply can be used as a concrete example. This is available in different units in the robot and gripping system model and must therefore be converted during transfer. If such knowledge is stored in knowledge graphs, the automated linking of models becomes possible.

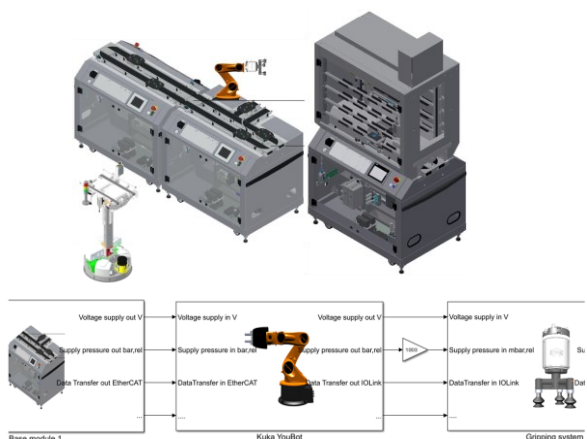

**Figure 2.** Simulation models and their coupling of inputs and outputs.

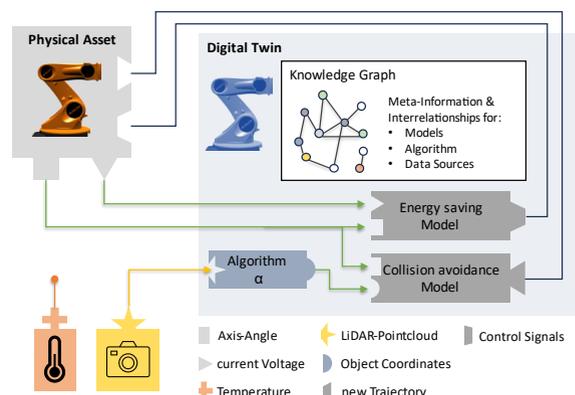

**Figure 1.** Enhancing asset control of a robotic manipulator via the Digital Twin using additional data sources.

The second use case pertains to utilize the Digital Twin during the operational phase. During this phase, simulation models of the Digital Twin are frequently executed in parallel with the corresponding asset. It is crucial for the Digital Twin to remain synchronized with the asset to give timely relevant insights. This necessitates a flow of information from the asset to the Digital Twin to minimize any deviations between asset and Digital Twin. Sensor data, which can originate from the asset or from its vicinity, is typically used for the synchronization of the Digital Twins' and assets' current state, or for gathering additional information useful in e.g., in simulation models of the Digital Twin. With the growing number of sensors and their connectivity capabilities, more data is available for such scenarios. However, assigning sensor data to corresponding inputs of a simulation model is becoming increasingly complex due to the large number of sensors. By describing the required inputs of the simulation models in Digital Twins and the information generated by sensors in a knowledge graph the manual mapping can be replaced with an automated mapping. This is exemplarily illustrated in Fig. 1, where the Digital Twin of a robotic manipulator uses the sensor data of the robot itself, but in addition enhances the own control with data generated from a LiDAR Sensor in the vicinity of the robot [7]. To automate the integration of such data, the model inputs, data transforming algorithms and available data sources need to be known and brought into relationship.

The third use case describes production scheduling in a discrete manufacturing factory. The schedule output of such a factory is to be optimized through various input scenarios using a material flow simulation. However, data from various sources is needed to set up the simulation model such as information about materials, processes, and machines. In addition, dynamic data, e.g. for orders, needs to be constantly updated and fed to the simulation model. This heterogeneous data landscape makes the creation and subsequent adaption of the model very time-consuming. A knowledge graph can be used to integrate the different data sources and provide a single access point to query the data. In addition, technologies like Shapes Constraint Language (SHACL) and Web Ontology Language (OWL) allow to validate data, so that the correct functioning of corresponding models is ensured [8]. Overall, this not only enables a simulation model to be generated and executed with fewer effort, but also offers the possibility of reflecting existing simulation results for model evaluation [9].

The use cases just described clearly show that semantic technologies can be used to benefit the Digital Twin in various ways. In addition, they show that semantic technologies can be useful in conjunction with simulation models in the Digital Twin. This results in the following research questions (RQ) for this contribution, from rough to fine:

- RQ 1: Are semantic technologies currently used in Digital Twins?
- RQ 2: For which tasks are semantic technologies used in Digital Twins?

- RQ 3: Are semantic technologies used in connection with simulation models in Digital Twins?

- RQ 4: How do semantic technologies support simulation models in Digital Twins?

The contribution is structured as follows. Section II explains the basic principles and definitions used in this work and highlights related studies that examine the use of semantic technologies in connection to Digital Twins or Simulation. Section III describes the methodology procedure of our review which is followed by the results of our analysis and the answering to the research questions in section IV. Our contribution is discussed in section V and finally concluded in section VI.

## II. BASICS AND RELATED WORK

This section provides the basic principles and a general overview of the paper. It introduces the concept of the Digital Twin and defines knowledge graphs and semantic technologies as they will be used throughout the paper. Finally, related literature reviews on Digital Twins and semantic technologies are presented.

### A. THE DIGITAL TWIN

The concept of the Digital Twin, which dates to its origins in the 2000s, is being used in a variety of fields, including healthcare, smart cities and manufacturing [10, 11]. Since 2016, there has been a significant increase in publications related to Digital Twins [12]. Over the years various definitions of the concept have emerged [13]. For this work, we will use the definition provided by [14], defining a Digital Twin as a virtual representation of a physical asset. Its main characteristics are simulation capability, synchronization, and active data acquisition. The core components of a Digital Twin are shown in light blue in Fig. 3. Active data acquisition, synchronization and executable models is particularly important for a Digital Twin. If Services, DT Model Comprehension, Intelligent Algorithms and a Feedback Interface are added, marked in dark blue, a Digital Twin can be extended to an intelligent Digital Twin [14]. The intelligence of the intelligent Digital Twin can manifest in various ways. For instance, it can automatically generate control code for newly added machines within the Plug & Produce framework, optimize the process sequence, or provide predictive capabilities using the operational data stored in the Digital Twin during runtime [4]. Knowledge graphs and semantic technologies can be an important enabler of intelligence in the digital twin, which can be used, for example, in the aspect of DT model comprehension or intelligent algorithms. The concept of the Digital Twin is widely used in the literature and in some cases includes aspects of the intelligent Digital Twin mentioned above. As the aim of this article is to conduct a structured literature review, the term Digital Twin, including the concept of the intelligent Digital Twin, will be used in the following sections.

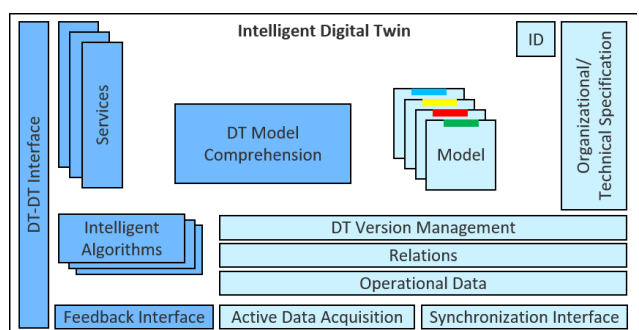

**Figure 3.** Concept of the intelligent Digital Twin [4].

### B. KNOWLEDGE GRAPHS AND SEMANTIC TECHNOLOGIES

The idea of representing knowledge in a semantically rich manner go back to 1950s, but only began to accelerate rapidly since the invention of the semantic web in the 2000s [15]. Since then, research focused on bringing knowledge representation to web standards, managing, and linking data, as well as creating potential applications for these. However, since a few years the concept of knowledge graphs began to gain big popularity focusing on knowledge representation within numerous applications while maintaining the large graph aspect [16]. Now with recent hype of large language models and generative artificial intelligence, knowledge graphs could gain another wave of popularity, as they contemplate the generic knowledge of large language models by giving them domain context and enhanced graph structure [17]. Nevertheless, many companies introduced their knowledge graphs and, although the internal implementation is different, the basic principles and characteristics always remain the same: The capability of combining diverse reasoning methods and knowledge representations while guaranteeing the required scalability, according to the reasoning task at hand [16, 18]. Paulheim defined the following four criteria of a knowledge graph [19]:

- It mainly describes real world entities and their interrelations, organized in a graph.

- It defines possible classes and relations of entities in a schema.

- It allows for potentially interrelating arbitrary entities with each other.

- It covers various topical domains.

However, a uniform definition for knowledge graphs remains still open. In the reminder of this paper, we will use the definition of [20]: A knowledge graph is "a graph of data intended to accumulate and convey knowledge of the real world, whose nodes represent entities of interest and whose edges represent relations between these entities". Due to this quality knowledge graphs offer more flexibility and adaptability than traditional technologies for data and knowledge representation. Semantic technologies on the other end go hand in hand with knowledge graphs. For the remainder of this paper, semantic technologies are defined a set of tools to derive or give meaning to data across applications and data sources. Knowledge graphs utilize a suitable subset of semantic technologies depending on the use case and purpose, such as schemas, inferencing, or logic. Thus, knowledge graphs can be seen as semantic technology solution. The World Wide Web Consortium (W3C) specified an extremely well-suited set of technologies for implementing semantics which evolves around Resource Description Framework (RDF) and is often referred to as Semantic Web Technologies. However, we don't want to restrict semantic technologies to the W3C standards.

### C. RELATED STUDIES
The Several surveys and reviews have been conducted in the past regarding Digital Twin and semantic technologies. The following section provides an overview of their contributions and covered topics.

In [21], the authors analyzed the two technology standards, Open Platform Communication Unified Architecture (OPC UA) and Automation Markup Language (AML), with respect to their semantic adaptability. The study shows that ontologies are often used as the basis for knowledge modelling and enable more flexible production orchestration. However, this review does not establish a clear connection between OPC UA, AML and the Digital Twin. [22] analyses the state of the art with respect to Digital Twins and identifies that several characteristics are shared among many Digital Twin publications. Although the analysis does not directly address semantic technologies, the authors acknowledge simulating the physical asset as one of the primary use cases in Digital Twins. This discovery, coupled with the identified research gaps in integrating semantically annotated data, highlights the importance of studying semantic technologies in the context of Digital Twins. [23] provides further insight into this topic. This systematic literature review (SLR) examines the enablers and barriers to the use of Digital Twins in the process industry. The findings may also hold relevance for the manufacturing domain. It is important to note that the authors of [23] primarily focus on the process industry. The study highlights system integration as a significant barrier due to the complexity of process plants. Furthermore, the authors identify knowledge building as a potential enabler for Digital Twins, highlighting the importance of analyzing knowledge graphs in Digital Twins. In their SLR the authors of [24]explore the combination of semantics and Digital Twins. While the review does not restrict itself to the industrial domain, most of the papers found are from this area. They analyze the use of ontologies in Digital Twins, identifying several use-cases. The article discusses the integration of heterogeneous data from multiple components, structuring of Digital Twin components, and reasoning-based relationship extraction. It identifies the usage of ontologies for describing physical and digital components and processes, mapping of communication protocols, and supporting the mapping of production processes with resources.

In summary, there already is a limited number on publications analyzing the literature regarding semantics and Digital Twins. Despite those available surveys, there is no study examining the correlation of knowledge graphs and Digital Twins with focus on the simulation aspect. Therefore, we aim to provide such a study.

### III. METHODOLOGY PROCEDURE OF THE REVIEW
To undertake a comprehensive review of the existing literature on the integration of knowledge graphs in Digital Twins, a systematic and rigorous methodology specifically for SLRs in software engineering areas was applied [25]. This section delineates the key steps involved in our SLR, aiming to provide transparency and reproducibility in the research process. The methodology is shown schematically in Fig. 4.

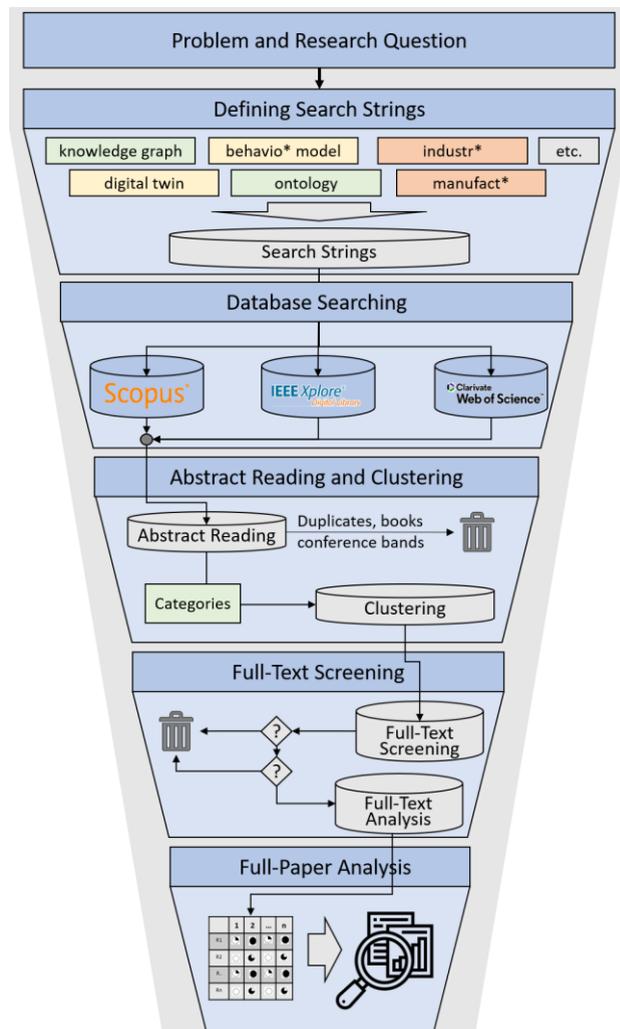

**Figure 4.** Methodology steps for the systematic literature review.

**Problem and Research Question:**

In an initial step, the problem must be defined, and the research questions derived in order to narrow down the scope for the literature analysis. Here, the problem to be investigated is the successful incorporation of knowledge graphs in the Digital Twin with a focus on simulation models. For a more detailed description of the problem, we refer to the use cases and research questions in section I.

**Defining Search Strings:**

In the first steps, a keyword collection is created based on the research questions, the authors' experience and a small literature research. This is done using generic terms such as Knowledge Graph, Digital Twin, Simulation, and Manufacturing. These terms are expanded with sub-terms and synonyms. To ensure the completeness and specificity of our literature search, various search strings and alternatives are formulated. These strings are developed to capture relevant publications dealing with the intersection of knowledge graphs, Digital Twins, and simulation models. The formulation of these search strings is done in an iterative process, combining the relevant keywords and phrases associated with the research questions. The used search strings are shown in TABLE I.

**Database Searching:**

The search for relevant literature is conducted across multiple databases chosen for their coverage of scientific and engineering research. The databases employed for this review includes Scopus, IEEE Xplore, and Web of Science. These platforms are chosen for their extensive coverage of scholarly articles, conference proceedings, and journals within the fields of computer science, engineering, and industrial technology.

**Abstract Reading and Clustering:**

Before the first screening process, the duplicates, books, and conference proceedings were sorted out. The abstracts were then read and analyzed. Keywords related to knowledge graphs, simulation models, and Digital Twins were identified. Based on these keywords, papers were clustered into distinct groups, facilitating a preliminary categorization of literature. The groups resemble the industrial domain, how well it fits to the definition of the Digital Twin, if semantics or knowledge is modeled, and finally if simulation is part of the Digital Twin. This step was essential in narrowing down the focus to papers that explicitly deal with both semantic knowledge bases and simulation models within the Digital Twin context. Based on the identified keywords and categories the publications are assigned for the Full-Text-Screening. Here publications containing keywords from manufacturing, mentioning simulation or Digital Twin as well as semantics, ontologies or knowledge are included.

**Full-Text Screening:**

Prior to the actual full-text screening, the availability of the papers was checked. If a paper was not available, it was excluded from the analysis. The screening then checked whether the content of the paper dealt with the explicit collaboration of knowledge graphs and digital twins or simulation in the field of manufacturing. The purpose of this classification is to exclude papers that only mention digital twins or simulation in the abstract or in the motivation, but do not deal with them conceptually or in terms of implementation.

**Full-Paper Analysis:**

The subsequent phase of our review involved a thorough examination of the selected papers. Each paper underwent detailed scrutiny, encompassing a comprehensive analysis of its methodologies, findings, and contributions. This full-paper analysis was critical for extracting nuanced insights, identifying trends, and evaluating the quality and relevance of the literature. A focus was maintained on discerning the methodologies employed by authors in integrating knowledge graphs within the Digital Twin framework.

By following this systematic approach, our literature review aims to provide a robust foundation for deriving essential components and insights for the successful implementation of knowledge graphs in Digital Twins. The methodological transparency is crucial for ensuring the reliability and validity of our findings, thereby contributing to the overall rigor of our research process.

TABLE I   USED SEARCH STRINGS DURING THE SYSTEMATIC LITERATURE ANALYSIS.

| Nr. | Search string |
|---|---|
| 1 | "Knowledge Graph" AND "Digital Twin" (Industr* OR Manufactur*) |
| 2 | "Knowledge Graph" AND ("Simulation" OR "Simulationmodel") AND (Industr* OR Manufactur*) |
| 3 | ("Ontology" OR "Semantic") AND ("Digital Twin" OR "Simulation") AND (Industr* OR Manufactur*) |
| 4 | "Knowledge" AND "Simulation" AND "Digital Twin" AND (Industr* OR Manufactur*) |
| 5 | "Knowledge Graph" AND "Simulation" AND "Parameter" AND (Industr* OR Manufactur*) |
| 6 | "Knowledge Graph" AND "Simulation" AND ("Input" OR "Output") AND (Industr* OR Manufactur*) |
| 7 | "Knowledge Graph" AND "Simulation" AND "modeling depth" AND (Industr* OR Manufactur*) |
| 8 | "Knowledge Graph" AND behavio* model AND "Digital Twin" AND (Industr* OR Manufactur*) |

TABLE II Quantity of Papers in the Individual Literature Analysis Steps

| Analysis Step | Number | Sort Out |
|---|---|---|
| Data Base Searching | 925 | 443 |
| Abstract Reading/Clustering | 482 | 305 |
| Full-Text Screening | 177 | 145 |
| Full-Paper Analysis | 32 | - |

## IV. LITERATURE ANALYSIS

In this section, we provide an overview of the results for the literature review. The quantitative results are presented first, followed by the answers to the RQs. Finally, we give an overview of our assessment of the analyzed works.

### A. QUANTITATIVE RESULTS OF LITERATURE ANALYSIS

The quantitative results of the individual steps of the literature analysis are summarized in TABLE II. In the first step of the method, a total of 925 results were collected from the three databases. In the second step, 186 conference proceedings, 2 books and 255 duplicates were sorted out. 482 contributions were subjected to abstract reading and clustering as described in Section III. Afterwards 305 publications got sorted out resulting in 177 articles subjected to full-text screening. As a result of the following classification step also described in Section III, a further 145 articles were considered as not relevant for answering the research questions in this contribution. As a result, 32 papers were evaluated in detail in the full-text analysis.

**Analysis of the Data Base Searching:**
In Fig. 5, the distribution of found literature is presented, segmented by the eight utilized search strings (see Table I). In addition to breaking down the search strings, the figure illustrates how many publications were found in the respective databases. Therefore, the diagram also gives an indication about the relative distribution of search results per data base. As can be seen, Scopus provided clearly the most literature results per search string, followed by Web of Science, whereas IEEE provided the least results. However, the representation has not yet been cleaned of duplicates, which could result in a big overlap of publications across databases and search strings. In addition, Scopus is the only database providing results for search strings 7 and 8. While IEEE could find titles related to search string 6, no results were obtained via Web of Science for strings 6, 7, and 8. This indicates the relevance of the Scopus database for the quantitative analysis of this research field. However, based on the available information, no qualitative assessment of the results can be made yet.

The distribution of Fig. 5 also reveals that most publications were found using search strings 3 and 4. This is likely attributed, in part, to their broader abstraction regarding the "knowledge-base" component of the search strings. The terms "semantic OR ontologies" and "knowledge," compared to "knowledge graph" in the other search terms, evidently include a significantly larger number of publications. This may suggest that the use of knowledge-based technologies in the Digital Twin is prevalent, whereas the exploration of knowledge graphs is not yet extensive. The fewest publications were obtained with search strings 5-8. This can be attributed to their additional specific formulations in areas such as simulation parameters, simulation inputs and outputs, modeling depth, and simulation modeling. Nevertheless, the extremely low number of available publications in this area indicates potential open research domains.

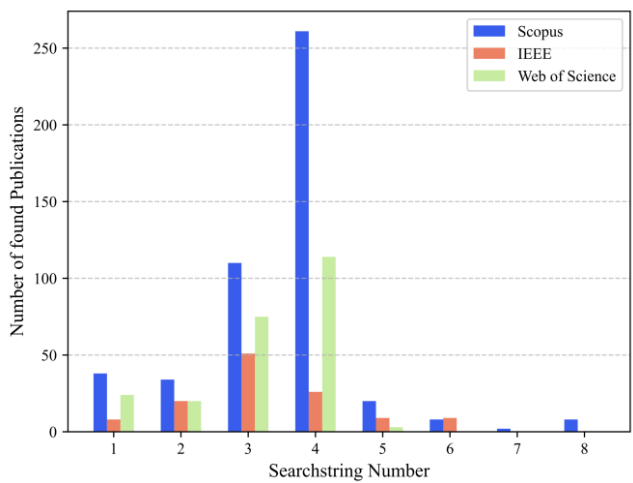

**Figure 5.** Distribution of the discovered publications, categorized by the respective databases.

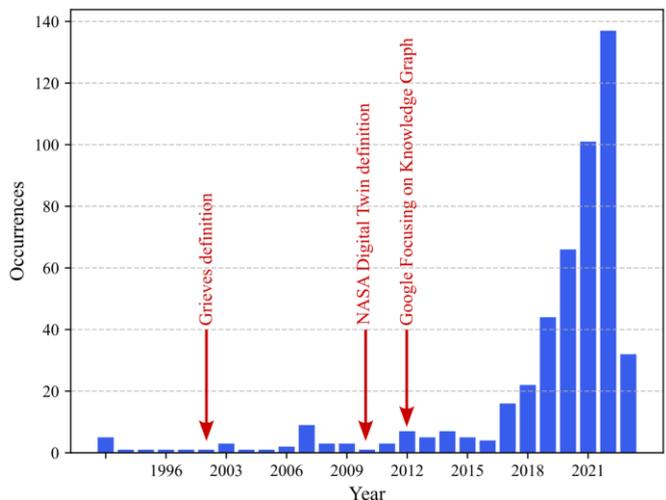

**Figure 6.** Distribution of publications per year across all search terms.

As evident from Fig. 6, there has been a significant surge in the number of publications, particularly in the last six years. Preceding this period, the field of combining Digital Twins and knowledge modeling seemed to undergo several waves, and the distribution of publications can be attributed to various aspects. As mentioned in Section "Digital Twin" was first mentioned in a company presentation by Michael Grieves in the early 2000s [26], primarily referring to a model for product life-cycle management. All publications before this date should thus address the combination of simulations and knowledge modeling. Based on the costs for computational resources back then, the usage of semantics in combination with simulation can be considered marginal. Using simulations in the engineering process of technical systems was already challenging. This might explain the barrier to complicate this topic further with knowledge modeling approaches. The first wave, where the combination of simulation techniques and knowledge modeling emerged, aligns closely with the introduction of the term "Digital Twin" in 2002 and the introduction of the semantic web technologies RDF and OWL in the early 2000s. However, this wave subsided. It was not until around 2010 that the second wave of relevant publications began. Around this time, the widely accepted definition of the Digital Twin by NASA emerged: "A Digital Twin is an integrated multiphysics, multiscale, probabilistic simulation of an as-built vehicle or system that uses the best available physical models, sensor updates, fleet history, etc., to mirror the life of its corresponding flying twin." [27]. Approximately simultaneously, Google introduced and promoted the concept of knowledge graphs [28]. These circumstances seem to reignite research in the field of the Digital Twin in combination with knowledge technologies. After the adoption of knowledge graphs and Digital Twins in their respective domain, the number of publications related to the combined research field began to increase around 2016/17. From that point onward, the quantity of identified publications increased rapidly.

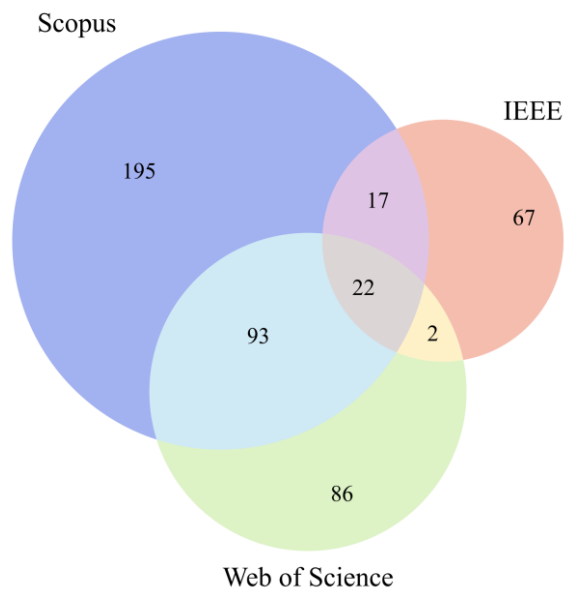

**Figure 8.** Overlap of the found publications between the used databases - Scopus, IEEE, and Web of Science.

Various factors may account for this. On the one hand, the quantity of available and interconnected machines and devices has also increased [23], leading to an exponential rise in available data [29]. Additionally, in the industrial sector, more applications are emerging based on a corresponding data foundation and knowledge modeling, such as Predictive Maintenance, Automatic Model Generation, etc.

**Analysis of the Abstract Reading and Clustering:**
Fig. 8 illustrates the distribution of publications found solely in one database versus those found in multiple databases. Conference proceedings, books, and duplicates have already been filtered out from this dataset. As evident in Fig. 5 and Fig. 8, the majority of publications were found in the Scopus database, followed by Web of Science and IEEE. The overlap between Scopus and Web of Science is substantial, with 93 titles, even exceeding the number of publications found exclusively in Web of Science. Both databases appear to have significant overlap in the areas of Digital Twin and Industry.

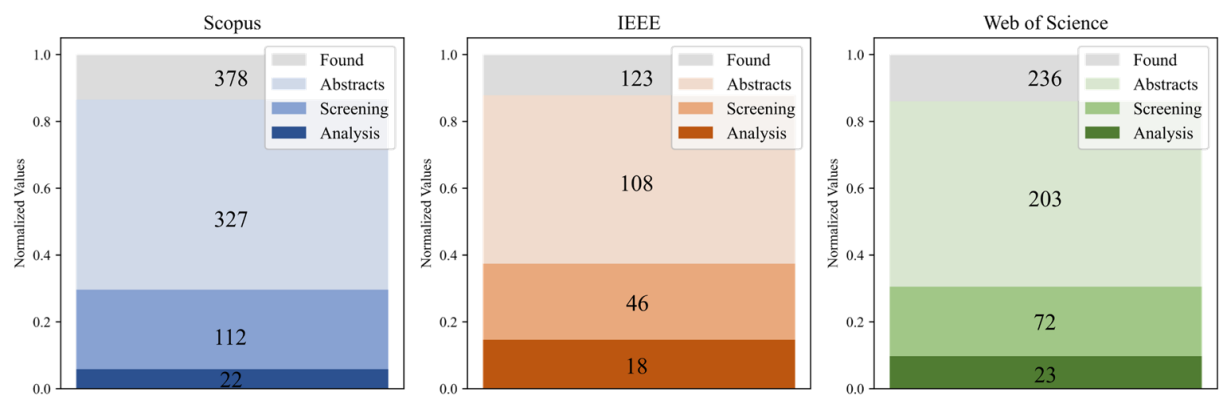

**Figure 7.** Number of papers found, used for abstract screening, full screening, and full analysis for the respective databases - Scopus, IEEE, and Web of Science.

This could be attributed to the broad focus of both databases, potentially resulting in a shallower pool of available literature in specific subject areas and a stronger emphasis on Open Access titles. The overlap for the IEEE database is considerably lower. Only 2 publications were found in both Web of Science and 17 in both Scopus and IEEE. This suggests that the IEEE database contains a significantly larger number of unique publications. This may be attributed to the technical specificity of the IEEE database. Additionally, events organized by IEEE (IEEE conferences) might contribute to the number of publications exclusively available in IEEE. In general, it can be concluded that integrating the IEEE database into a literature search in the technical field leads to a significant improvement in the discovery of relevant literature.

**Analysis of the Full-Text Screening:**
In Fig. 7, the three bar plots represent the number of papers used in the corresponding filtering steps during the SLR Process. On the y-Axis of each plot the normalized value in respect to the amount of found papers is shown. In the middle of the bar-plot the absolute number of papers for each bar is shown. The bar plot is not stacked, meaning every bar is starting from y = 0.0 %. As can be seen in the absolute numbers, although the absolute number of papers is significantly higher for *Scopus* than e.g. *IEEE*, the percentage of papers passing further filtering steps for *IEEE* is larger (*Scopus* = 6.7 %, *IEEE* = 16.6 %, *Web of Science* = 11.3 %). This implies that the quality of found papers respective the factor of fitting to the search terms is about three times higher than it is in Scopus – an important factor when selecting databases for technical SLRs.

## B. FINDINGS REGARDING Q1 AND Q2
RQ1 and RQ2 are about the current use of semantic technologies and the tasks they currently solve within the Digital Twin (Section I). To answer these questions, we created a classification schema to assess the degree of usage of semantic technologies as well as clusters for the use case of semantic technologies in the Digital Twin. The classification schema for the usage of semantic technologies can be seen in Fig. 9.

| | |
|---|---|
| Deductive Knowledge | Reasoning, Logic, Rules |
| Querying | SPARQL, Cipher |
| Schema | RDFS, OWL, SKOS, Shacl, ShEx |
| Models | RDF, LPG |
| Identity | Unicode, URIs, IRIs |

**Figure 9.** Classification schema for semantic technologies.

As can be seen, we distinguish between 5 classes: *Identity, Models, Schema, Querying,* and *Deductive Knowledge.* These classes are derived from the well-known semantic web tech stack picture, as e.g. used in [30] and are generalized for different knowledge graph and semantic technologies to not only include the W3C technologies which are based on RDF. The names of the classes were partially derived from the introductory paper about knowledge graphs from [20]. On the ground level we have *Identity*, which is a key aspect of semantic technologies to ensure an entity can be identified globally by means of unique identifier like IRIs, URIs, or identity links. The second class is *Models*, which is the formalism in which the data and semantics are captured. Examples for this are RDF and Labeled Property Graphs, but also other standardized models that are not building knowledge graphs can be used here as e.g. ECl@ss. These two first classes are considered the minimum if semantic technologies are used in a standardized manner and consequently if *Models* are used properly, *Identity* is fulfilled as well. The third class is *Schema*, which can be implemented very differently but in general means to use a high-level structure or semantics that the data or graph follows. RDF, OWL, and SKOS are examples for a semantic vocabulary, that can be used to describe the data, whereas SHACL and ShEx are schemas used to validate if the data is following the defined schema. The constraints that are representing the validation schema are usually referred to as shapes. The fourth class is *Querying* which also includes the graph pattern on which the query languages are based. Examples are SPARQL for RDF graphs, or Cipher for LPG. The last class is *Deductive Knowledge* which includes mechanism to induce new data by means of a process, that was built up a-priori. This includes e.g. automatically inferring new data and the relationships between the data from the used data schema, which is referred to as reasoning. Also predefined rules or description logic which generate new data fall in this class.

The findings of our literature review regarding the used semantic technologies can be depicted from Fig. 10. It shows the distribution of the use of semantic technologies for the analyzed paper. Note that from the 32 publications that we analyzed, 5 were from the same authors and about the same topic so that we merged them for the analysis which results in 27 publications for the following graphs. As can be seen, more than 80% of the paper analyzed, use semantic technologies in a standardized manner. Interestingly, 2 works used standardized models, which in this case are industry standards, but do not incorporate them to semantic standards, which leads up in the smaller number for the use of *Identity*. Nearly three quarters of the contributions used *Schema* on top of that which signifies the need for a proper semantic description of the production domain as part of the Digital Twin. In most of the cases, this was realized in creating an OWL ontology to represent the data. Only 2 paper that were using *Models* and *Identity* didn't use *Schema*, but in addition used *Querying* which indicates that data was connected in a semantic manner and formalized in a standard way, but the actual meaning was not that important for the use cases. One thing to point out, is that the actual retrieving of the data as part of the Digital Twin only played a role in less than half of the works analyzed. Which leads to the direction that the use of knowledge graphs in Digital Twins is still not developed, and other technologies for accessing data as part of the Digital Twin are used. Three publications used the combination of *Identity*, *Models*, *Schema* and *Deductive Knowledge* and only another three times the whole semantic technology stack was used, indicating that the use case is highly responsible for the semantic technologies used and most of the approaches are not going for a full-fledged semantic representation but mostly use them for data representation.

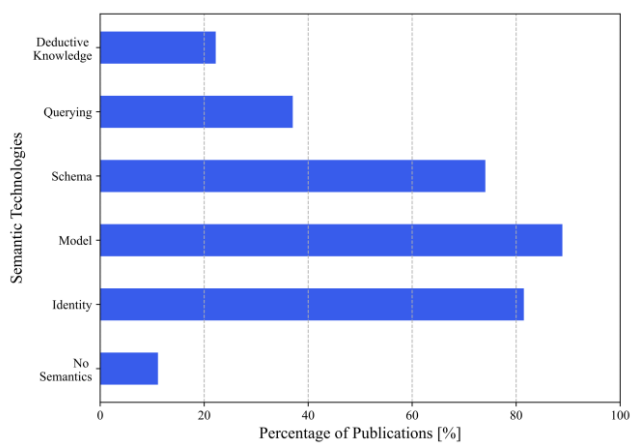

**Figure 10.** Classification schema for semantic technologies.

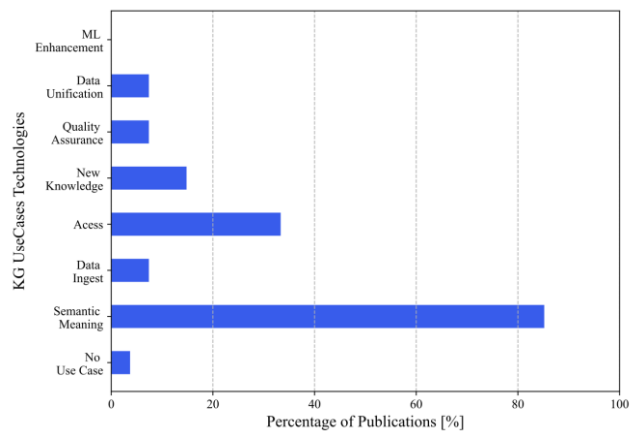

**Figure 11.** Applications for the use of Knowledge Graphs and semantic technologies.

This, however, leads to answering RQ2, about the use of semantic technologies within the concept of the Digital Twin. For this we clustered the use cases as follows:

- *Data unification/Interoperability*: Different data and sources need to be unified, mapped to each other and eventually also integrated in the same system.

- *Access/Availability:* Typically, heterogenous data is made available and accessible through one entry point (e.g. SPARQL endpoint), by means of data integration, or data virtualization.

- *Semantic Meaning/ Data Representation:* Data is described in a unified vocabulary so that the meaning and interpretation of the data is clear to all users (machine and human).

- *Data Ingestion:* Data is added to an already existing data set and put into context to this data set, typically in a very dynamic and high frequent way.

- *Derive new knowledge:* Deriving new knowledge from already existing one or from a data schema or a combination of both.

- *Quality Assurance:* Assure the quality of the data by means of constraints, visualization, or schema.

- *ML Enhancement:* Enhancing ML and AI with semantic interrelationships between the data that is fed to ML/AI, e.g. through graph embeddings.

The distribution of the use cases for the analyzed works can be depicted from Fig. 11. As can be seen, the primary use of semantic technologies in Digital Twins is *Semantic Meaning*. The big majority of the paper has this at least as a combined goal and similarly one can see that the technology to support this is mostly the *Schema*, *Identity*, *Models* cluster. This means, that currently the use of semantic technologies is still very limited and does not extend to more sophisticated technologies as reasoning or validation. Another big cluster seems to be the *Access* and *Availability*, which points to the direction that Digital Twins should be able to access data over a broad scope. However, *Data Ingestion* doesn't play a big role, indicating that the continuous update and synchronization of the Digital Twin with dynamic data is either not part of the concept of Digital Twins yet or not done with the help of knowledge graphs and semantic technologies.

## C. FINDINGS REGARDING Q3 AND Q4
RQ3 and RQ4 are about the use of semantic technologies in connection to simulation models and how they support simulation as part of the Digital Twin. To answer RQ3 we, access the works in regard to their ability to give semantics to simulation specific data. We differentiate between three different cluster for this:

- *Simulation Input Context:* Production data needs to be interpreted in the context of the simulation model and tool. This context and the link to the input production data can be semantically described.

- *Simulation Experiment Data:* This includes all data that is specific for the simulation experiments and necessary for the execution of the simulation model but is not directly given by the input of the production data, as e.g. user-specific parameters, like execution time.

- *Simulation Output:* When executed, the simulation model will generate data. This includes, state-specific data, like simulation logs but also the output and results of the simulation model.

- *Simulation Meta-Information:* All other simulation specific data or data about the simulation model are included in this cluster. Examples are information about the interconnection to other models, about the requirements and use of the simulation model.

Our findings can be seen in Fig. 12. As can be seen only 10 of the 27 contributions that we analyzed in depth, are using any form of semantic models to capture simulation specific information. Most of the papers are capturing the *Simulation Input Context*, which seems reasonable as in the context of Digital Twins a simulation model needs to be fed by production data. However, only two papers are going to that extent that they are describing every model aspect semantically, which might indicate that for most use cases this is simply not necessary and therefore not implemented. Surprisingly, *Simulation Meta-Information* ranks second behind *Simulation Input Context*. A reason for this is, that also works that are just considering the *Simulation Output* and the interconnection between simulation models, are included here. Nevertheless, the minority of paper is considering simulation semantics at all. This may hint to the fact, that for the current approaches of Digital Twins, no interconnection of multiple models is considered yet, but the predominant challenge is the describing of the simulation models in context with the used data. The context, however, seems to be currently described in the program code of the Digital Twin and is currently not explicitly represented, as may not be necessary for the current use cases of Digital Twins.

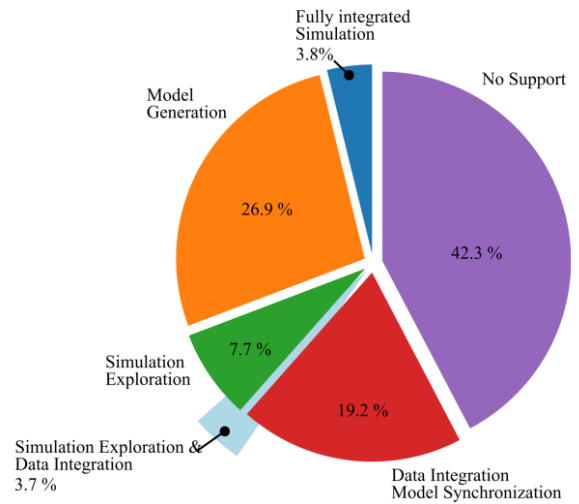

**Figure 13.** Classification schema for semantic technologies.

To answer RQ4, we created a cluster for the kind of simulation support that is provided. Here we distinguish between the following:

- *No support*: No support to simulation is given.

- *Data Integration/Model Synchronization*: The data of the production system is semi- or fully-automatically integrated to the simulation models in the Digital Twin. This can be implemented very differently but at least requires some form of semantic understanding of the data and an integration service that is doing the data transfer to the respective simulation components. *Model Synchronization* goes beyond *Data Integration* and is the capability to keep the Digital Twin aligned to the actual production system. Consequently, *Model Synchronization* needs a mechanism to deal with outdated data.

- *Model Generation*: Similarly, to the *Data Integration*, data is somehow described and mapped to simulation components. However, *Model Generation* does not only involve integrating one specific kind of data to be integrated (e.g. all dynamic data) but includes at least all data that is necessary to generate an executable simulation model. Again, the implementation can look very differently, e.g. the data can be mapped to predefined building blocks, or existing simulation components of a simulation tool and then be built up by a service.

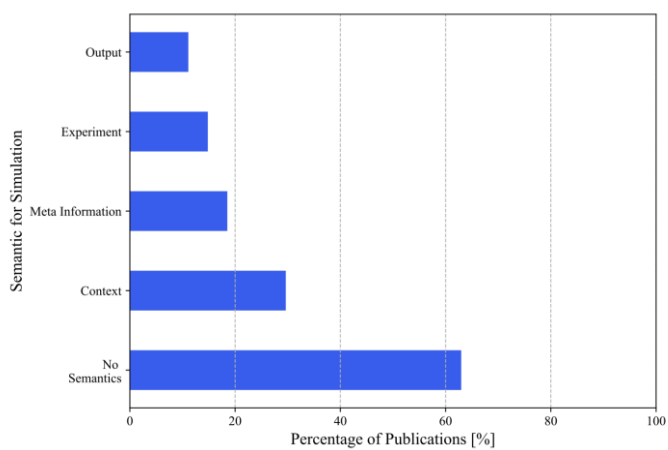

**Figure 12.** Semantics uses in Simulation Context

TABLE III OVERVIEW OF THE ASSESSMENT CRITERIA TO ANSWER THE RESEARCH QUESTIONS

| Publication | RQ's assessment criteria | | | | |
|---|---|---|---|---|---|
| | KG in iDTs (overall) | Data Capturing | Degree of Semantics | Simulation Semantics | Standardization |
| P. Vrba et al. 2003; 2012 [31,32] | ◐ | ◐ | ◐ | ○ | ◐ |
| Poudel et al. 2022 [33] | ○ | ○ | ◐ | ○ | ◐ |
| Rožanec et al. 2020; 2022 [34,35] | ◐ | ◐ | ◐ | ○ | ● |
| T. Wagner et al. 2007; 2014 [36,37] | ◐ | ◐ | ● | ○ | ● |
| Okamoto et al. 2021[38] | ○ | ○ | ◐ | ◐ | ◐ |
| Block et al. 2018 [39] | ○ | ◐ | ○ | ○ | ◐ |
| J. H. Lee et al. 2012 [40] | ◐ | ○ | ● | ○ | ◐ |
| J. Zhang et al. 2018 [41] | ○ | ◐ | ◐ | ○ | ◐ |
| Kadar et al. 2013 [42] | ◐ | ◐ | ◐ | ○ | ● |
| Kübler et al. 2020 [43] | ○ | ● | ○ | ◐ | ○ |
| Tsinarakis and Tsinaraki 2013 [44] | ○ | ○ | ◐ | ○ | ◐ |
| Wang et al. 2021a [45] | ○ | ◐ | ○ | ○ | ◐ |
| Zehnder and Riemer 2019 [46] | ◐ | ◐ | ◐ | ○ | ◐ |
| Zheng et al. [47] | ● | ○ | ● | ◐ | ● |
| K. Vernickel et al. 2022 [48] | ◐ | ◐ | ◐ | ○ | ◐ |
| P. Novák and R. Šindelář 2013 [49] | ◐ | ● | ◐ | ◐ | ◐ |
| Sahlab et al. 2021; 2022; Mueller et al. 2022 [50-52] | ◐ | ● | ◐ | ○ | ● |
| Schmidt and Pawletta 2013 [53] | ○ | ○ | ○ | ○ | ○ |
| Wang et al. 2021b [54] | ◐ | ● | ◐ | ○ | ◐ |
| Listl et al. 2022 [55] | ● | ● | ◐ | ● | ● |
| Hu et al. 2022 [56] | ◐ | ◐ | ◐ | ○ | ● |
| Jurasky et al. 2021 [57] | ● | ◐ | ◐ | ◐ | ◐ |
| May et al. 2022 [58] | ● | ● | ● | ● | ◐ |
| Mohammed et al. 2022 [59] | ◐ | ◐ | ● | ○ | ◐ |
| Schilberg and Meisen 2009 [60] | ○ | ○ | ○ | ◐ | ○ |
| Steinmetz et al. 2018 [61] | ○ | ○ | ◐ | ○ | ◐ |
| Zhao et al. 2021 [62] | ● | ● | ◐ | ◐ | ◐ |

- *Simulation Exploration*: *Simulation Exploration* is the task of simplifying the exploration and understanding of the simulation model. Semantic models are especially good suited for this task as they give a semantic meaning to the underlying data. Consequently, it does not involve any type of data transfer or mapping but only the data representation.

- *Fully integrated Simulation*: Here, the whole data management of the simulation is done within semantic technologies. This includes all production data as well as every data generated from the simulation model (output, parameters, log data from the simulation run).

Fig. 13 shows the distribution of the simulation support with the help of semantic technologies. From this we can conclude that the primary focus of the contributions is to bring production data to simulation models, either in the combination with services that create an executable simulation model or just the integration of data to the model. However, one can see that the simulation support is not considered for nearly half of the works, which shows that the simulation specific aspects of the Digital Twin is not yet fully established and interconnected. This is also supported by the observation that simulation semantics are not highly considered yet. Also, only one paper is taking the approach of doing a complete data management within semantic technologies.

## D. FINDINGS OVERVIEW

To summarize our findings about the use of semantic technologies in the Digital Twin, we created a harvey's ball diagram for the analyzed paper which can be seen in Table III [31–62]. To assess the literature, we formed denser cluster for the research questions RQ1 and RQ3, that were used in the previous section. We did not include RQ2 and RQ4 in the table as these research questions are about the use of semantic technologies in the Digital Twin, and such a gradation of the cluster was not properly displayable. For RQ1, an empty ball indicates basic semantics, like *Identity* and a *Models*, a half ball indicates a bit more advanced toolset of semantic technologies like *Schemas* and *Querying*, and finally a full ball indicating strong use of the semantic technologies like *Deductive Knowledge* and a mix of different technologies. For RQ3, we assessed the degree of semantic technologies for simulation specific aspects, which means an empty ball is considering no simulation semantics, a half ball is representing a partwise consideration of simulation semantics, and a full ball is capturing almost all parts of the simulation in semantic technologies.

In addition to the RQs, we assessed the works for two more criteria. First, we considered the data that was captured in the knowledge graph, as the major strength of them is to integrate heterogenous data and it also gives an indication about the difficulty of using semantic technologies and creating the knowledge graph. We differentiated between three degrees of heterogeneity: homogeny (empty ball), medium heterogeny (half ball) and heterogeny (full ball). Homogeny means that primarily static production data from one source or a few sources has been captured as e.g., equipment or product data. Medium heterogeneity indicates that static and dynamic data was captured from a few sources, as e.g. orders and execution data. Heterogeny indicates that in addition to production data also other information, like environmental data, simulation data or meta-data was represented and used from a very heterogenous data landscape. The second criteria that we assessed was if industry standards were used for the semantic technologies as this gives an indication about the reusability of the approach. This includes standards for the domain knowledge which would be implemented in the semantic schemas (e.g., ISA-95, OPC UA, ...) as well as standards for the semantic technologies (OWL, RDF, Cipher, ...). Here, an empty ball indicates no use of standards at all, a half ball indicates either standards in domain knowledge or semantic technologies, and a full ball represents the use of both. Finally, we included an overall assessment how sophisticated semantic technologies are used in the Digital Twin in regard to simulation so far, which is represented in the first column of Table III.

As can be seen no publication achieves to cover all aspects of the semantic technologies in order to be part of the Digital Twin, although [55] and [58] just come short on one aspect. Interestingly, there seems to be a discrepancy between the data capturing and the degree of semantics. Only one paper achieves to have apply a high degree of semantics to a wide scope of data, which might indicate that even though highly promoted as strength, it can be quite difficult to apply semantic technologies properly in a heterogenous data world. Standardization is used quite frequently throughout all the observed work, and even used seven times in both semantic technologies and domain knowledge. However, most of the works that were assessed with a half ball, were only implementing semantic technologies in a standardized manner, and were lacking domain standardization. Finally, we can see again, that almost all the contributions are not considering simulation semantic in a very sophisticated manner, indicating that the research field of the Digital Twin is not yet heavily exploring on the possible use cases that go along with that as e.g. model interconnection and, model data management. The current research, however, seems to be focusing more on the representation of the data that is needed by the simulation.

## V. DISCUSSION
Knowledge graphs and semantic technologies in general are promoted highly as potential enhancers and vital elements of the concept of the Digital Twin [14, 63]. In fact, when taking a look at the concept of the Digital Twin of section II and the interconnection between both technologies, we see they can support each other in multiple areas, namely the feedback interface, active data acquisition, synchronization interface, and DT Model Comprehension. As also evident from the use case descriptions in section I, semantic technologies and knowledge graphs usually enable the following tasks in this context:

- Semantic description of data, including the used production data and generated data from simulation models in the Digital Twin.

- Semantic description of models in the Digital Twin (simulation, AI, ...), including the relationship of the models to the data as well as connections to other models. This gives context to the model in the Digital Twin on the one hand and context to the data regarding the model, e.g. by describing the input, output, and parameters on the other hand.

- Dealing with dynamic data and data changes, and consequently putting context to the new data within the old data. This is necessary to know how the data is used inside the models and which data can be neglected.

In addition, more advanced methods like reasoning, and validation constraints can be used to check consistency and correctness of the data. Queries can be used as universal access for services inside the Digital Twin (e.g. for the models to enable simulatability) and outside the Digital Twin (e.g. for model understanding, or manual updates).

However, as can be seen in the results of the systematic literature review, implementations of the Digital Twin do not cover such aspects yet. Most of the contributions examined, are only covering one aspect of semantic technologies and only very few are dealing with model descriptions at all.

Nevertheless, we see the advantage of the standardized and unified representation of the models, that can be used to create standardized APIs to access the Digital Twin and connect it to different applications. By that, the semantics in Digital Twin enhance scalability and widen the scope of potential applications for the Digital Twin such as mentioned use cases in section I. Specifically, the continuous integration and update of data context of the used models (simulation, AI) is necessary to enable a synchronized representation and simulation of the production system.

## VI. CONCLUSION

Digital Twins and semantic technologies have received a lot of attention in the literature in recent years. By integrating semantic technologies into the Digital Twin, new types of use cases can be realized. This applies in particular to the property of simulatability and the associated use cases of Digital Twins. This contribution examines existing approaches from this combination with a focus on simulation. A total of 925 papers were evaluated in a systematic literature review, 32 of which were analyzed in detail. The following key aspects are highlighted:

- None of the papers covers a comprehensive approach that realizes the support of simulation in the Digital Twin, but often only partial aspects.

- The combination of Digital Twins and knowledge graphs is an ongoing research activity, which shows the relevance of these topics, however, it seems still very difficult to combine both on a large scale.

- The degree of semantics is currently highly dependent on the use case the Digital Twin is implemented for. Consequently, the adaptability of the simulation models of the Digital Twin to different use cases is currently limited.

- Simulation over the entire lifecycle and its efficient creation and use is becoming increasingly important and expected.

- With their semantic information provision, knowledge graphs provide a good basis for supporting Digital Twins in the future with autonomous decisions efficiently and confidentially.

It has been shown that there are many sub-concepts for different use cases. Combining these approaches into a comprehensive approach can create new synergies in the context of the Digital Twin. In addition to the challenges of bringing these approaches together technologically, standardization work is also required here. Innovative approaches such as LLMs models can then efficiently enable a wide variety of use cases on the basis of this information provision. For security-related and confidential results, however, research into well-founded knowledge bases is essential.

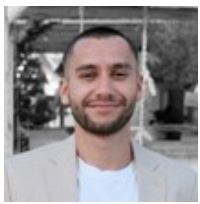

**Franz Georg Listl** is a PhD student at the Institute of Industrial Automation and Software Engineering (IAS) at the University of Stuttgart, Germany and Research Scientist at Siemens Technology in Munich, Germany. He received his bachelor's degree in mathematics from the OTH Regensburg in 2017 and completed his master's degree in the same field in 2019. Since 2020 he is pursuing his Ph.D. under the supervision of Prof. Dr.-Ing. Michael Weyrich with Siemens Technology. Since 2023, he has been a research scientist at Siemens Technology. His research interests include digital twins, simulation, and knowledge graphs in production and automation. He is particularly interested in the use of knowledge graphs and semantic technologies for material flow simulation.

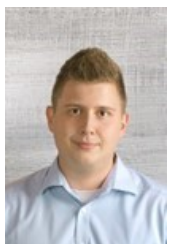

**Gary Hildebrandt** was born in Leonberg, Baden-Württemberg, DE in 1995. He received his B.Eng. in Mechanical Engineering in 2019 from Pforzheim University and completed his M.S. in Mechatronic Systems Engineering in 2021, also at Pforzheim University. Since 2021 he is pursuing his Ph.D. in Automation Engineering at Pforzheim University in cooperation with Stuttgart University. He is the Author of multiple Articles and conference Publications. His research interests include Digital Twins for the manufacturing industry and engineering for modular production systems. Additionally, he is exploring current possibilities in the field of mixed-reality engineering, where he combines content from real and virtual production environments to ease the engineering of future production systems.

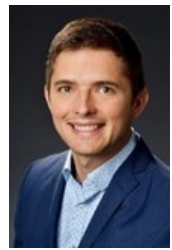

**Daniel Dittler** is a PhD student and research assistant at the Institute of Industrial Automation and Software Engineering (IAS) at the University of Stuttgart, Germany. He received his bachelor's degree in mechanical engineering from Pforzheim University in 2019 and his master's degree in mechatronic system development in 2021. His areas of interest are digital twins in automation technology and the automatic model adaption of behavior models in the operational phase.

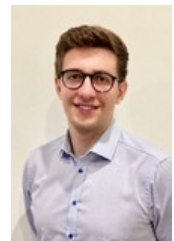

**Valentin Stegmaier** is a PhD student at the Graduate School of Excellence advanced Manufacturing Engineering GSaME at the University of Stuttgart, Germany. He received his bachelor's degree in electrical engineering and information technology from the University of Stuttgart University in 2017 and completed his master's degree in the same field in 2020, also at the University of Stuttgart. Since 2020 he is pursuing his Ph.D. under the supervision of Prof. Dr.-Ing. Michael Weyrich with J. Schmalz GmbH. His areas of interest are digital twins in automation technology with focus on the efficient creation of behavior models in process relevant modeling depths.

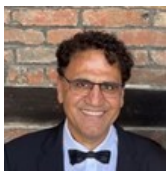

**Dr.-Ing. Nasser Jazdi** received his Diploma in Electrical Engineering in 1997, his Ph.D. (Topic: Remote Diagnosis and Maintenance of Embedded Systems) in 2003 from the University of Stuttgart, Germany. In 2003, he joined the Institute of Industrial Automation and Software Engineering (IAS), University of Stuttgart as Deputy Director, researcher and lecturer. Since 2019, he has been lecturing AI for Industrial Automation at the University of Anhui, VR China. His research interests include Dynamic Calculation of Reliability using Machine Learning, Software Reliability in the context of IoT and artificial intelligence in industrial automation. He is an IEEE senior member, a member of the VDE (The Association of German Engineers) and a member of VDI-GPP (Company Product and process design) Software Reliability Group. He is the author of 15 journals and 75 conference publications. He also has three book chapters. Dr. Jazdi is the Deputy Director of the Institute of Industrial Automation and Software Engineering at the University of Stuttgart and visiting Professor at the Anhui University in Hefei.

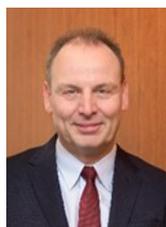

**Prof. Dr-Ing. Michael Weyrich** studied electrical engineering, specializing at the University of Applied Science Saarbrücken, University of Westminster (London, U.K.) and at the Ruhr-University Bochum, focussing on automation technology. Thereafter, he worked as a Research Assistant with Prof. Dr.-Ing. Paul Drews of the European Centre of Mechatronics and was awarded a doctoral degree in 1999 from the RWTH Aachen. Prof. Weyrich worked with Daimler AG for eight years, where he assumed the leadership of the "End-to-End Process Flexible Manufacturing" project and later had a technical management function in the "CAx Process Chain – Production" of the Information Technology Management division. From 2004 onward, he was the head of "IT for Engineering" at DaimlerChrysler Research and Technology Bangalore (India). After his return to Germany, he joined Siemens AG as department leader with direct report to the Business Unit Head of Motion Control in Erlangen for 2 years, focusing on innovative technologies. In 2009, he was appointed as Professor for the Chair of Automation in Manufacturing at the University of Siegen and as Director of the Automotive Centre Südwestfalen. Prof. Weyrich assumed the role of Director of the Institute of Industrial Automation and Software Engineering at the University of Stuttgart in 2013. He is very interested in research in cyber physical systems for industrial application. He has published over 100 papers and is very active in research and is a member of the Board of the German Engineering foundations VDI/VDE GMA. Additionally, he is an appointed reviewer for the European Commission, the German Research Foundation, DFG and a number of other institutions.